\documentstyle[12pt]{article}
\def\o{\over}
\def\A{\rightarrow}
\def\bar{\overline}
\def\r{\gamma}

\def\a{\alpha}
\def\b{\beta}
\def\n{\nu}
\def\m{\mu}

\def\e{\epsilon}

\def\th{\theta}

\def\bar{\overline}

\def\G{{\rm GeV}}

\def\eV{{\rm eV}}
\def\be{\begin{equation}}
\def\ee{\end{equation}}
\def\bea{\begin{eqnarray}}
\def\eea{\end{eqnarray}}
\def\lsim{\mathrel{\lower2.5pt\vbox{\lineskip=0pt\baselineskip=0pt
           \hbox{$<$}\hbox{$\sim$}}}}
\def\gsim{\mathrel{\lower2.5pt\vbox{\lineskip=0pt\baselineskip=0pt
           \hbox{$>$}\hbox{$\sim$}}}}
\begin{document}
\setlength{\baselineskip}{8mm}
\begin{titlepage}
\begin{flushright}
\begin{tabular}{c c}
& {\normalsize  EHU-96-4} \\
& {\normalsize \today}
\end{tabular}
\end{flushright}
\vspace{2 mm}
\begin{center}
{\large  \bf Is Zee Model The Model of Neutrino Masses?} \par
\vspace{8 mm} 
Alexei Yu. Smirnov $^{a,}$ \footnote{E-mail:
\ smirnov@ictp.trieste.it}, \quad
Morimitsu Tanimoto$^{b,}$\footnote{E-mail:\
tanimoto@edserv.ed.ehime-u.ac.jp}   \\ 
{\it 

${}^a $ International Center for Theoretical Physics \\
     I-34100 Trieste, Italy \\
${}^b $Science Education Laboratory, Ehime University, \\
      Matsuyama 790, JAPAN \\
}
\end{center}

\vspace{7 mm}


\begin{abstract} 

Zee model leads naturally to 
two heavy, strongly degenerate and 
almost maximally mixed neutrinos and to one light neutrino  
with  small mixing. 
This pattern  coincides  with the one needed  for 
a solution of  the atmospheric neutrino problem 
by $\n_\m\A \n_\tau$ oscillations 
and for  existence of the two component hot dark matter 
in the Universe. 
Furthermore, the oscillations 
$\bar\n_\m\A \bar\n_e$ 
can be in the range of sensitivity of  
KARMEN, LSND experiments.   
Phenomenology of this 
scenario is considered and  possibility to check it 
in the forthcoming experiments is discussed. 
Scenario  
implies large values and inverse flavour  hierarchy of the couplings 
of the Zee boson with fermions: 
$f_{e \tau} \ll f_{\mu \tau} \leq f_{e \mu} \sim 0.1$. 
Main signatures of scenario  are: strongly suppressed 
signal of $\n_\m\A \n_\tau$ oscillation  
in CHORUS and NOMAD experiments,  
so that positive result 
from these experiments will  rule out the scenario;  
possibility of observation of 
$\n_e\A \n_\tau$ oscillations by CHORUS and NOMAD; 
corrections to the 
muon decay, neutrino-electron scattering at the level 
of the experimental errors;    
branching ratio $B(\mu \A e \gamma)$  bigger than $10^{-13}$. 
The solar neutrino problem can be solved 
by introduction of additional very light singlet fermion  
without appreciable changes of the 
active neutrino pattern.\\  

\end{abstract}
\end{titlepage}

\section{Introduction}

 Zee model \cite{zee}  is the simplest model which explains a smallness of 
neutrino masses by physics at the electroweak scale. 
 It can be considered as an alternative of the see-saw 
mechanism \cite{see}. 

Zee model 
suggests the existence of charged scalar field
$h$, singlet  of the $SU(2)$, and  two  
doublets of the Higgs bosons $\Phi_1$, $\Phi_2$.
The singlet  $h$  couples to lepton doublets
$\Psi_{lL}=(\n_l, l^-)$, ($ l = e, \mu, \tau$) 
as well as to Higgs doublets; 
leptons are assumed to couple to 
doublet $\Phi_1$ only. The appropriate terms in the Lagrangian 
are     
\begin{equation}        
{\cal L}_{Zee} = 
f_{\ell \ell'} \Psi_{\ell L}^T i\tau_2  \Psi_{\ell'L} h \ 
+ c_{12} \Phi_{1}^T i\tau_2 \Phi_{2} h^\dagger  
+ \frac{m_l}{\langle \Phi_1 \rangle} \bar \Psi_l \Phi_1 l_R  
+ h.c. , 
\label{lagr}
\end{equation}
\noindent
where $c_{12}=-c_{21}$ are real mass parameters,  
the couplings  $f_{\ell \ell'}$  are 
antisymmetric in $\ell$ and $\ell'$. 
The interactions (\ref{lagr}) generate  neutrino mass terms in one 
loop. 
 
Zee model gives very distinctive pattern of  
neutrino masses and mixing.  
For not too strong hierarchy of the couplings $f_{\ell \ell'}$ 
the two heavy neutrinos,  
$\nu_2$,  $\nu_3$,  are strongly 
degenerate  and mix almost maximally in $\nu_{\mu}$,   
$\nu_{\tau}$. The first neutrino $\nu_1$ practically coincides  
with $\nu_e$ and has much smaller mass: 
\be
m_1 \ll m_2 \approx m_3 . 
\label{hierarchy}
\ee 
It was marked \cite{stao} 
that this pattern  
coincides with the one needed to 
solve  simultaneously the  atmospheric neutrino problem \cite{atm} 
and the problem of the hot dark matter in the Universe \cite{hdm}. 
Indeed, the deficit of the atmospheric muon neutrinos 
can be explained by the oscillations $\nu_{\mu} - \nu_{\tau}$ 
with practically maximal mixing. Two heavy 
neutrinos with masses 
$m_2 \approx m_3 \approx (1 - 5)$ eV compose two component hot 
dark matter (which may give even better fit of the 
cosmological data than 
one component) \cite{hdm}. Furthermore, the oscillations 
$\nu_{\mu} \A \nu_e$ 
and $\bar\n_\m\A \bar\n_e$  
can be at the level of sensitivity 
of existing experiments: BNL \cite{E776},  
 KARMEN \cite{KAL}  (see \cite{stao}). Later 
it was marked \cite{wolf} \cite{anj} that the  model can 
immediately accommodate positive LSND  result \cite{LSND}. \\ 
 
In this paper we will consider phenomenology  
of the outlined scenario, and in particular,  
the possibility to check it  
by forthcoming experiments.  
In sect. 2 we describe the scenario in  details. 
Sect. 3 is devoted to oscillations. In sect. 4 we 
find the bounds on the Zee coupling constants. 
In sect. 5 implications of  data on the muon decay,      
neutrino electron scattering, $e- \mu - \tau$ universality 
to the scenario are considered.      
Predictions for  $\mu\A e\r$ and $\nu_{3(2)}\A \nu_1 \r$ are given. 
In sect. 6, we describe a modification  
of the Zee model which is  able to solve the 
solar neutrino problem. Sect. 7 contains our  conclusions.

\vskip 0.2 cm

\section{\bf  Scenario}
\par  
The  neutrino mass matrix of the Zee model 
in flavor basis, $\nu = (\nu_e, \nu_{\mu}, \nu_{\tau})$,  
can be written as 
\begin{equation}
     M =  m_0 \left ( \matrix {   0 & \e &  \sin\th \cr
      \e & 0 & \cos\th \cr
 \sin\th & \cos\th & 0   \cr } \right ) \ , 
\label{matr}
\end{equation}
\noindent
where $m_0$ is the basic mass scale.   
Mixing angle $\theta$ and  parameter 
$\e$  can be naturally  much smaller than 1.   
(We will discuss the relation 
of these parameters with the parameters of 
the Lagrangian (\ref{lagr}) in sect. 4.) 

In the case $\cos\theta \gg \sin \theta, \e$ 
the eigenvalues of matrix (\ref{matr}) are
\begin{equation}        
   m_1 = -m_0 \e  \sin 2\th \ , \qquad   m_{2,3} = m_0 (\pm 1 - {1\o 2} \e
\sin 2\th )\ ,
\label{eigenval}
\end{equation}
\noindent
and the mixing matrix $S$ which 
diagonalizes (\ref{matr})  is 
\begin{equation}
\label{mmatr}       
 S \simeq  {1\o \sqrt{2}} \left (
     \matrix { \sqrt{2}\cos\th & \sin\th + \e\cos\th &  
\sin\th - \e\cos\th \cr
-\sqrt{2}\sin\th &  \cos\th & \cos\th \cr
  -\sqrt{2}\e  & 1 & -1  \cr } \right )  \ ,
\end{equation}
($\nu_f = S\nu$,  where   $\nu\equiv (\nu_1,\nu_2,\nu_3)$ 
are the mass eigenstates).  
According to (\ref{eigenval}) 
the states $\nu_2$ and  $\nu_3$ are approximately degenerate, and
their masses ($\sim m_0$) are much larger than the mass of $\nu_1$.
The mass squared difference is
\begin{equation}        
   \Delta m_{32}^2 = 2\e \sin 2\th m_0^2 \ll m_0^2  \  ,
\label{relation}
\end{equation}
\noindent
 where 
   $\Delta m^2_{ij} \equiv m^2_i-m^2_j$.
For $\nu_1$ component model gives  
 $\Delta m_{21}^2\simeq \Delta m_{31}^2\simeq m_0^2$ and 
  the ratio of mass differences equals  
\begin{equation}        
   {\Delta m_{32}^2 \o \Delta m_{21}^2 } = 2\e \sin 2\th  \  .
 \end{equation}
 \noindent 
 Thus, the Zee mass matrix  gives two different scales
 for the mass squared differences and  the maximal mixing between 
 two heaviest neutrinos.
        \par
 
As it was outlined in the introduction we will consider the 
following scenario: 

\begin{itemize}

\item
$\nu_2$ and $\nu_3$ form the two component 
hot dark matter so that 
\be
m_0 = m_{HDM} =  (1 - 5) ~ {\rm eV}. 
\label{HDM}
\ee

\item 
The $\nu_{\mu} - \nu_{\tau}$ 
oscillations with practically 
maximal depth explain  the atmospheric neutrino deficit 
and therefore 
\be
\Delta m_{32}^2 = \Delta m_{atm}^2 \sim  (0.3 - 3) \cdot 10^{-2} 
~{\rm eV^2} .     
\label{atm}
\ee

\item 
The oscillations 
$\nu_{\mu} - \nu_{e}$ and 
$\bar{\nu}_{\mu} - \bar{\nu}_{e}$ 
with $\Delta m_{21}^2 \approx m_{HDM}^2$ 
can be  in the region of sensitivity 
of the KARMEN and LSND experiments. For 
$m_0$ in the cosmologically interesting 
domain (\ref{HDM}) this means that 
\be
\sin^2 2\theta \leq    
\sin^2 2\theta_{e\mu} \sim 
(1 - 3) \cdot 10^{-3} \ ,
\label{theta}
\ee
where $\theta_{e\mu}$ is the experimental bound 
(or preferable value in the case of  positive result) 
for the $\nu_e - \nu_{\mu}$ mixing angle. 

\end{itemize}

Substituting $m_0$ and $\sin^2 2\theta$ from 
(\ref{HDM}) and (\ref{theta}) in  (\ref{relation}) 
we get 
\be
\e \geq  \frac{\Delta m_{atm}^2}
{2 m_{HDM}^2 \sin 2\th_{e\mu}}  \  . 
\label{epsilon}
\ee
According to   (\ref{atm}) and
 (\ref{epsilon}) $\e = 10^{-3} - 0.5$, 
with typical value $3\cdot 10^{-2}$. 
Thus all oscillation parameters of the model 
(mass squared differences and mixing angles) 
can be  fixed by the experimental data immediately. 


\section{\bf  Neutrino Oscillations}

In terms of the elements of the  
mixing matrix (\ref{mmatr}),  
$S_{\a i}$, the oscillation probability 
can be written as 
\begin{equation}        
P(\n_\a\A \n_\b) = \delta_{\a\b} - 4 \sum_{i>j} 
               S_{\a i} S_{\b i} S_{\a j} S_{\b j} 
            \sin^2\left({\Delta m^2_{ij} L \o 4 E_\n}\right) \ , 
\label{oscill}
\end{equation}
\noindent  
where $E_\n$ is the neutrino energy and 
$L$ is the distance.  
We neglect the $CP$ 
violation, so that the elements $S_{\a i}$ are real.  
Let us consider  the probabilities (\ref{oscill})
for  short and long distances 
separately. 
 
1). In the short distance limit   
the phase difference due to
$\Delta m^2_{32}$ is small:  
${\Delta m^2_{32} L \o 4 E_\n} \ll 1$.  
Taking into account that 
$\Delta m^2_{31} \simeq \Delta m^2_{21}$ 
and using matrix (\ref{mmatr}) we find 
\begin{equation}        
P(\n_\m\A \n_e) \approx  
P(\bar{\n}_\m\A \bar{\n}_e) 
= \sin^2 2\th \sin^2 \left({\Delta m^2_{31} L \o 4 E_\n}\right) \ .
\label{emu}
\end{equation}
This result is applied to E776 \cite{E776}, 
KARMEN \cite{KAL} and  
LSND \cite{LSND}.

For $\nu_{\mu} - \nu_{\tau}$ we get 
\begin{equation}        
P(\n_\m\A \n_\tau) = 
4 (\e \sin\th)^2 \sin^2\left({\Delta m^2_{31} L \o 4 E_\n}\right) 
+ \cos^2\th \sin^2\left({\Delta m^2_{32} L \o 4 E_\n}\right) \ .
\label{mutau}
\end{equation} 
The $\nu_{\mu} - \nu_{\tau}$  
oscillations with large mass splitting are doubly 
suppressed because of $\sin\th\ll 1$ and  $\e\ll 1$.
This smallness is related to the fact that according to 
(\ref{mmatr}) the admixture  of $\nu_1$ in $\nu_{\tau}$ 
is suppressed by $\e$. The mode 
of oscillations with the smallest mass splitting 
(second term in (\ref{mutau})) may give a comparable 
contribution. For values of $\sin^2 2\theta$ and 
$\e$ from  (\ref{theta}) 
and  (\ref{epsilon}) correspondingly   
we obtain 
\begin{equation} 
P(\n_\m\A \n_\tau)\sim  10^{-7} - 10^{-5} \ .
\end{equation} 
If both  $\e$  and $\Delta m^2$ are near the upper bounds 
the probability can be as big as $10^{-4}$.
These values  are still below the 
sensitivity of  CHORUS and NOMAD \cite{CHONOM}, 
but they may be in the regions of sensitivity of 
planning experiments  E803 at Fermilab and
E889 at BNL \cite{E803889}.

For $\nu_e - \nu_{\tau}$ channel we get 
\begin{equation}        
P(\n_e\A \n_\tau) = 
4 (\e \cos\th)^2  \sin^2 \left({\Delta m^2_{31} L \o 4 
E_\n}\right)  \ . 
\label{etau}
\end{equation}
\noindent
If $\e \geq 10^{-1}$ and $m_0 > 4$ eV,     
then $P(\n_e\A \n_\tau) \simeq 10^{-2} - 10^{-1}$ 
and  the $\n_e\A \n_\tau$ oscillations   
can be detected by 
CHORUS  and NOMAD.    

Thus  the observation  
of signals of the $\n_e\A \n_\tau$
oscillation and absence of signal from  
$\nu_{\mu} \A \n_\tau$ mode in
CHORUS and NOMAD  are the signatures of the Zee 
model. The scenario under consideration  will 
be ruled out if  CHORUS and NOMAD
find  signals of the $\n_\m\A \n_\tau$ oscillations.

According to (\ref{relation}) for fixed 
$\Delta m_{32}^2$ and $m_0$, 
the parameter $\e$  is  inversely proportional to $\sin \th $.
Therefore  $P(\n_e\A \n_\tau)$ increases when  $P(\n_\m\A \n_e)$ decreases,
as is shown in Fig.1.         
\begin{center}
\unitlength=0.7 cm
\begin{picture}(2.5,1.5)
\thicklines
\put(0,0){\framebox(3,1){\bf Fig. 1 }}
\end{picture}
\end{center}
In particular, if  $P(\bar \n_\m\A \bar \n_e) \sim 3\cdot 10^{-3}$ 
(the level of the LSND result)  
and $m_0^2 > 6$ eV$^2$,  
then $P(\n_e \A \n_\tau) <  3 \cdot 10^{-5}$  
which is beyond the sensitivity of the  CHORUS and NOMAD.
On the contrary, for 
$P(\n_e \A \n_\tau) > 2 \cdot 10^{-2}$ which can be observed 
by these experiments one has 
$P(\n_\m\A \n_e) < 10^{-5}$. 
Thus a comparison of  results from searches for 
$P(\n_e \A \n_\tau)$ and  
$P(\n_e \A \n_\mu)$ oscillations can  
give crucial check of the model.

For the parameters under consideration 
there are  strong resonance transitions  
$\n_e \rightarrow \n_\mu, \n_\tau$ and  
$\n_\mu \rightarrow \n_e$ 
in the inner parts of the collapsing stars. 
As the consequences one predicts: 
(i) disappearance of the neutronization peak, (ii) 
hard $\n_e$ spectrum  at the cooling stage, 
(iii) additional energy release in the 
inner parts of star which will stimulate shock 
wave revival desired for the star explosion. 
(iv) In the same time the 
$\n_\mu \rightarrow \n_e$ conversion  
leads to  suppression of the r-processes 
responsible for nucleosynthesis 
of heavy elements unless $m_0\leq 2\eV$ 
\cite{full}.

2). In the long distance limit  experiments are sensitive to 
oscillations stipulated by small mass difference 
$\Delta m_{32}^2$ and the oscillations due to 
large mass  difference are averaged out. 
We get the results   
\bea 
 P(\n_\m\A \n_e) &=& {1\o 2}(\sin 2\th)^2
  - \cos^2\th (\sin^2\th-\e^2 \cos^2\th)
\sin^2({\Delta m^2_{32} L \o 4E_\n}) \ , 
    \\
  P(\n_\m\A \n_\tau) &=& 2 (\e \sin\th)^2 
     + \cos^2\th \sin^2\left({\Delta m^2_{32} L \o 4 E_\n}\right) \ , 
\eea
which are applied to the atmospheric neutrinos.  
Notice  that $P(\n_\m\A \n_e)$ is suppressed due to both 
$\sin\th\ll 1$ and  $\e\ll 1$,  
and the dominant effect comes from 
$\n_\m\A \n_\tau$ oscillations as we suggested in the introduction. 
\vskip 0.5 cm
\noindent

\section{\bf Parameters of Zee Model}\par  
  
In terms of parameters of the 
Lagrangian (\ref{lagr}) 
the elements of the mass matrix (\ref{matr}) 
equal  
\begin{equation}  
\tan\th\equiv {f_{e\tau}\o f_{\mu \tau}} \ ,  \ \ \ \  
\e \equiv  {f_{e\mu}\o 
\sqrt{f_{e\tau}^2 +  f_{\mu \tau}^2}} 
\left(\frac{m_{\mu}}{m_{\tau}}\right)^2 ,  
\label{thetae}
\end{equation}  
and  \cite{zee} \cite{pet} 
\begin{equation}   
m_0 \approx m_{\tau}^2 
\sqrt{f_{e\tau}^2 +  f_{\mu \tau}^2} 
{g \sin 2\phi \cot\b \o
64 \sqrt{2} M_W  \pi^2} \ln{M_{2}^2\o M_{1}^2} \ .
\label{mzero}
\end{equation}
Here $m_{\tau}$ is the tau lepton mass,  
$g$ is the weak coupling,  
$m_W$ is the $W$-boson mass,   
$\tan\beta \equiv <\Phi_1>_0/<\Phi_2>_0 $ is the ratio 
of the VEV of two Higgs doublets.  The angle $\beta$ 
determines physical charged  Higgs boson: 
$\Phi^+=\Phi_1^+ \cos\b - \Phi_2^+ \sin\b$, where
$\Phi_1^+, \ \Phi_2^+$ are two charged Higgs fields from the  doublets. 
The angle 
$\phi$ is the mixing angle of the Zee singlet 
and the physical charged component of the Higgs doublet  
$\Phi^+$:   
%
\begin{eqnarray}        
   h &=&\cos\phi H_1 + \sin \phi H_2 \ , \nonumber\\
   \Phi^+ &=& -\sin\phi H_1 + \cos \phi H_2 \ , 
\end{eqnarray}   
\noindent 
where $H_1$ and $H_2$ are the   eigenstates 
of the mass matrix with masses $M_{1}$ and $M_{2}$,   
and the mixing angle is determined by 
\begin{equation}        
\tan 2\phi = {4\sqrt{2} g^{-1} c_{12} M_W \o \sqrt{(M_{1}^2-M_{2}^2)^2
                  - (4\sqrt{2} g^{-1} c_{12} M_W)^2}}  \ .
\end{equation}

As we have seen in  sect. 2,  the parameters of the mass matrix 
(\ref{matr}) $m_0$, $\epsilon$, $\theta$ can be  fixed by the 
data. This in turn allows one to 
determine  the ratios of the constants $f_{ij}$ 
using (\ref{thetae})    
\be
\frac{f_{e \tau}}{f_{\mu \tau}} = \tan \theta_{e\mu} \ll 1 \ ,
\label{ratio}
\ee
\noindent and
\be
\frac{f_{e \mu}}{f_{\mu \tau}} \approx   
\frac{\Delta m_{atm}^2}{2 m_{HDM}^2}\cdot
\left(\frac{m_{\tau}}{m_{\mu}}\right)^2 \cdot
\frac{1}{\sin 2\theta_{e\mu}} \ .
\label{ratio2}
\ee
For $\sin^2 2\theta_{e\mu} =2\times 10^{-3}$,  
$\Delta m_{HDM}^2=6 \ \eV^2 $ and
 $\Delta m_{atm}^2=10^{-2} \eV^2 $ 
Eq. (\ref{ratio2}) gives 
$f_{e \mu}/f_{\mu \tau} = 5.3$ 
which means an inverse hierarchy of the 
Zee boson couplings  with $f_{e \mu}$ being the 
largest one \cite{anj}. 
For fixed value  $P(\n_\m\A \n_e)$ the mixing angle 
$\theta$ is the function of 
$\Delta m_{31}^2 = m_0^2$. 
Using this dependence we get from 
(\ref{ratio}) and  
(\ref{ratio2})  
the ratios $f_{e \mu}/ f_{e \tau}$ and $ f_{\mu \tau}/ f_{e
\tau}$ as the functions of $m_0$   
for  fixed value of $P(\n_\m\A \n_e)$ (see Fig.2).   
For 
$P(\n_\m\A \n_e) = 1.5\times 10^{-3}$, 
(which is in the range of sensitivity of KARMEN and LSND)   
we find $f_{e \mu} \simeq f_{\mu \tau} \gg f_{e \tau}$
at  $m_0=5$ eV. 
This relation may testify for  certain  horizontal
symmetry.  
Below $m_0=5$ eV, there is an  inverse flavour 
hierarchy of the couplings, 
     $f_{e \mu} \geq f_{\mu \tau}  \gg f_{e \tau}$. 
For  $P(\n_\m\A \n_e) \leq  10^{-4}$
  one gets the inverse flavor hierarchy already below
$m_0=10$ eV.\\ 

\begin{center}
\unitlength=0.7 cm
\begin{picture}(2.5,1.5)
\thicklines
\put(0,0){\framebox(3,1){\bf Fig. 2}}
\end{picture}
\end{center}
The absolute value of the coupling constants can be fixed by 
(\ref{mzero}). 
For  values of parameters:
$\sin\phi\simeq O(10^{-1})$, $\tan\b\simeq O(10)$, $M_{1}\simeq 
M_{2}\simeq O(500\G)$ we get: 
$f_{e\mu} = 10^{-2} - 1$. That is the scenario 
implies quite big couplings constants of the Zee boson. 

\vskip 0.5 cm

\section{\bf Constraints 
from  the Electroweak Processes}\par

Since the  constants 
$f_{e\mu}$, $f_{\mu \tau}$ are rather big the Zee singlet 
can  give observable  contributions to different weak processes. 
The effective four-fermion Lagrangian induced by the 
Zee boson exchange can be written (after appropriate 
Fiertz transformation) as 
\be
 \frac{G_F}{\sqrt{2}} \xi  
\left[ 
\bar \n_\m \gamma^{\mu} (1 - \gamma_{5}) e 
\bar \n_\m \gamma_{\mu} (1 - \gamma_{5}) e -  
\bar \n_\m \gamma^{\mu} (1 - \gamma_{5}) \mu 
\bar \n_e \gamma^{\mu} (1 - \gamma_{5}) e + ...\right] \ ,  
\label{hamilton}
\ee 
where 
\be
\xi \equiv \left(\frac{1}{\sqrt {2} G_F} 
\frac{f^2_{e\mu}}{\bar M^2} \right) \ , 
\label{Z}
\ee
and 
\begin{equation} 
{1\o \bar M_{H}^2} \equiv 
{\cos^2\phi\o M_{1}^2}+{\sin^2\phi\o M_{2}^2} \  . 
\label{mbar}
\end{equation} 
Notice that  only usual left handed components of leptons 
participate in the interactions with Zee boson, 
and therefore the Lagrangian (\ref{hamilton}) has usual $V - A$ form. 

In the case  neutrino electron scatterings, 
$\bar \n_\m \  e^- \A \bar \n_\m \  e^-$ and
$\n_\m \ e^- \A  \n_\m \ e^-$ , 
the contribution from  (\ref{hamilton}) leads to 
a change of the $g_L^e$ coupling: 
$g_L^e \A g_L^e + \xi$.  CHARM II experimental data 
on $g_L$ and $g_R$ \cite{charm}  agree well 
with predictions of the Standard Model.   
Therefore  $\xi$ should be 
smaller than the experimental error 
$\Delta g_L^e$: $\xi < \Delta g_L^e$.    
Using (\ref{Z}) we have explicitly
\be
\frac{f^2_{e\mu}}{\bar M^2} < 0.036 G_F \ . 
\label{Zbound}
\ee 
The Zee singlet exchange leads also to the lepton number 
violating    
process $\n_\m \ e^- \A  \n_{\tau}  \ e^-$ 
which contributes 
to $\nu_{\mu} e$ scattering incoherently. Its amplitude 
is proportional to $f_{e\mu} f_{e\tau}$.

The second term  in the Lagrangian (\ref{hamilton}) 
gives  the 
renormalization of the four fermion coupling $G_{F}$ 
of the muon decay. Assuming that  the 
effect of the  Zee 
boson  on the decay rate is smaller than $0.1\%$, 
(so that it does not destroy the agreement in the electroweak
precision tests) we find   
\be
\frac{f^2_{e\mu}}{\bar M^2} < 7 \cdot 10^{-4} G_F \ . 
\label{mubound}
\ee 
Also the modes of the muon decay with lepton number violation appear: 
$\mu \rightarrow \nu_{\tau} \ e \ \bar\nu_e$, 
$\mu \rightarrow \nu_{\mu} \ e \ \bar\nu_{\tau}$, 
$\mu \rightarrow \nu_{\tau} \ e \ \bar\nu_{\mu}$ which  
 contribute to the total decay rate incoherently. 

The result (\ref{mubound}) allows one to get the  bounds on masses 
and mixing of 
scalar bosons. Indeed, using expression for the mass 
(\ref{mzero}) we can find 
$f_{e \mu}$ as the function 
of $\phi$, $\beta$ and $M_i$. Substituting  
$f_{e \mu} = f_{e \mu}(\phi, \beta, M_i)$
into (\ref{mubound}) we find the lower bound on 
$\sin \phi$ as the function of $M_1$ for different values 
of $P(\n_\m\A \n_e)$,   
$m_0$, $\tan \beta$  
and fixed  $M_2=300 \ \G$ (see Fig.3 (a)$-$(c)). 
Notice that the most strong bound is for 
$M_1 = M_2$. Forbidden region becomes larger 
with increase of $\tan \beta$ as well as with 
decrease of $m_0$ and  $P(\n_\m\A \n_e)$.    
Big  region 
of parameters exists in which all the restrictions are satisfied. 
Furthermore, one of the charged Higgses can be at the level 
of lower kinematical 
bound.

\begin{center}
\unitlength=0.7 cm
\begin{picture}(2.5,1.5)
\thicklines
\put(-2.5,0){\framebox(8,1){\bf Fig. 3 (a), 3 (b), 3 (c)}}
\end{picture}
\end{center}

The bound on the model follows also from 
$e - \mu - \tau$ universality. 
The expected deviation from universality due to the Zee boson
contribution is 
$|1 - g_\tau/g_\m| \sim  (f_{e \mu}^2)/(G_F \bar M^2)$,  
where $g_{\mu}$ and $g_{\tau}$ are the weak coupling constants 
of the charged currents with $\mu$ and $\tau$. 
 Recent measurement of the 
branching ratio of the decay $\tau\A e\bar\n_{e} \n_{\tau}$
  at OPAL \cite{opal} gives  
the ratio of couplings  
$g_\tau/g_\m = 1.0025\pm 0.0060$,  
and the corresponding  bound on the parameters of model 
is weaker than (\ref{mubound}).\\


The  model leads to the radiative decays of the  
muon  $\mu \A e\r$  
and neutrino $\nu_{3(2)}\A \nu_1 \r$  through the one-loop 
diagram with Zee singlet.  
  
The branching ratio of the $\mu \A e\r$ \cite{zee} \cite{pet} is 
 \begin{equation}   
  B(\mu \A e\r) = \left({\alpha \o 48 \pi }\right)
     \left ( {f_{e\tau} f_{\m \tau}\o \bar M_H^2 G_F} \right)^2  \ .
  \end{equation}
Using (\ref{thetae}), (\ref{mzero}) and  (\ref{mbar}) 
we can express  it as  
$B(\mu \A e\r) = A(\sin\phi, M_{i}, \tan \beta) m_0^4 $. 
The   branching ratio becomes smaller  
with increase of  $\sin\phi$ 
and decrease of $\tan\b$ (see fig. 4).   
The present experimental upper bound $B < 4.9\times 10^{-11}$ \cite{PDG} 
(shown  by the horizontal dashed line)   
will be strengthen soon up to $5\times 10^{-13}$ 
by the experiment at MEGA in LAMPF(Los Alamos).  
Future  experiment   \cite{kuno} will 
push   the limit to  $3\times 10^{-14}$.  
The results from these  experiments  
combined with bounds from precision tests (fig.3) will 
cover essential part of the parameter space of the model.\\ 
\begin{center}
\unitlength=0.7 cm
\begin{picture}(2.5,1.5)
\thicklines
\put(0,0){\framebox(3,1){\bf Fig. 4}}
\end{picture}
\end{center}
 
The life time of $\nu_i\A \nu_1 \r(i=2,3)$ 
equals \cite{zee} \cite{pet}:
  \begin{equation}   
    \tau(\nu_{i}\A \nu_1) = 
\left \{ \a m_i^5  \left [ 2{m_\mu^2\o m_\tau^2}
C_\mu   \left (1-{C_\tau\o C_\mu} \cos 2 \th \right ) \right ]^2 
           \left (1-{m_1\o m_i}\right )^3  \right \} ^{-1} \ ,
\label{decay}  
\end{equation}
  \noindent  where
  \begin{equation} 
   C_\ell ={1\o \ln \left ({M_{H2}^2\o M_{H1}^2}\right )} 
    \left [  
\frac{1}{M_{H2}^2} 
\left(\ln \left ({M_{H2}^2\o m_\ell^2} \right )-1 \right)  
- (2 \A 1)\right ] \ ,
        \quad \ell=\m, \tau \ .
\label{c}
\end{equation} 
The life time $\tau(\nu_{i}\A \nu_1)$
 depends mainly on the charged Higgs scalar masses $M_{1}$ and 
  $M_{2}$;  $f_{\ell \ell'}$  and $\sin\phi$ 
enter only via the mass of neutrino.                  
For  $m_0=1  - 10 \ \eV$ 
the  life time is  in the interval 
$10^{22} - 10^{29}$ years. This may have some cosmological 
implications.   
 
In the limit of $f_{e\tau}=0$  
the anomalous magnetic moment of neutrino which 
corresponds to (\ref{decay})
equals  \cite{pet} 
\begin{equation}   
\m_{\nu}
          \simeq - 4 e m_0 C_{\tau} \ ,
  \end{equation}
where $C_{\tau}$ is defined in (\ref{c}). 
For  $M_{1}\simeq M_{2}\simeq 300 \ \G$ and $m_0 = 2.65 \  \eV$, 
we get $\m_{\nu} \simeq 6 \times 10^{-16} e/2m_e$.\\

\section{\bf Solar neutrinos}

For solar neutrinos  all 
oscillations are averaged and from (\ref{oscill}) 
one gets survival probability 
\be
P(\nu_e \rightarrow \nu_e) = 
\cos^4 \theta + \frac{1}{2} \sin^4 \theta + O(\e^2) \ .
\label{survival}
\ee
There is no dependence of  suppression 
of the $\nu_e$ - flux on energy and for 
$\e, \sin^2 \theta \ll 1$ the effect is small. Thus in the 
considered scenario there is no 
solution of the solar neutrino problem. 

Let us suggest that apart from three known neutrinos 
also singlet (because of the LEP bound) 
neutrino $\nu_s$ exists. This neutrino mixes 
with electron neutrino so that the resonance 
conversion $\nu_e \A \nu_s$ explains the deficit of the 
solar $\nu_e$-flux. The explanation requires 
the mass squared difference and 
the mixing angle in the intervals \cite{solar} \cite{MSW}: 
\be
\Delta m^2 = (4 - 10) \cdot 10^{-6}  {\rm eV}^2 \ , 
 \qquad \sin^2 2\theta_{es} = 10^{-3} - 10^{-2}. 
\label{parameters}
\ee
The singlet neutrino could be the  right handed 
counterpart of the known neutrino components or 
new very light fermion which comes from 
some other sector of theory.

The mass of the lightest neutrino in the Zee model 
(which is essentially the $\nu_e$) is 
\be
m_1 = m_0 \e \sin 2 \theta \approx 
\frac{\Delta m^2_{atm}}{2 m_{HDM}} 
\sim (1 - 5) \cdot 10^{-3} {\rm eV} \ .  
\label{mass1}
\ee
Squared mass  $m_1^2$ is close to $\Delta m^2$ 
desired for solar neutrinos (\ref{parameters}). 
This means that the mass of singlet neutrino, $m_s$, 
should be rather close to $m_1$ (recall that for 
the resonance conversion one needs $m_s > m_1$): 
\be 
\frac{m_s - m_1}{m_1} \approx \frac{\Delta m^2}{2 m_1^2} \ . 
\ee
For $m_1 > 4 \cdot 10^{-3}$ eV one gets from this equation 
$\Delta m/m_1 < 0.2$ .

Let us consider the simplest scheme with only 
one  singlet neutrino. We extend the Lagrangian 
of the Zee model by adding the  terms: 
\be
f_l \bar \Psi_l \Phi \nu_s + m_{ss} \nu_s^T \nu_s  \ . 
\label{slagran}
\ee
All couplings $f_i$ can be  of the same order. 
The first term leads to  mixing of the $\nu_s$ with 
the active neutrinos: $m_{ls} = f_l \langle \Phi \rangle$. 
Performing block diagonalization of $4\times 4$ mass 
matrix we get the mass matrix for 
the $(\nu_s - \nu_e)$ system: 
\be
M \approx 
\left(
\begin{array}{ll}
m_{ss} & m_{es}\\ 
m_{es} & m_1 
\end{array}
\right) , 
\ee
where $m_{es} = f_e \langle \Phi \rangle$
and $m_1$ is fixed in (\ref{mass1}). 
The mixing angle is then 
\be
\sin 2\theta_{es} \approx \frac{2 m_{es}}{m_{ss} - m_1} \ .
\ee
If the mass $m_{ss}$ is not too close to $m_1$,  we get 
\be
m_s \sim (2 - 4)\cdot 10^{-3} {\rm eV} \ , \qquad  \ m_{es} \sim 10^{-4} 
{\rm eV}. 
\label{ms}
\ee
With increase of $m_s$ (and consequently the degeneracy) 
$m_{es}$ can be further diminished.  

According to (\ref{ms}) a solution of the solar neutrino problem implies 
very small Yukawa coupling $f_{e} < 10^{-15}$ which is 
of the order  $m_{EW}/ m_{string}$, 
where $m_{string} \sim 10^{18}$ GeV  is the superstring scale. 
The mass of the singlet neutrino $m_s$ 
is of the order $m_{3/2}^2 /m_{string}$ which also 
may indicate the SUSY origin of the singlet. 

Mixing of the singlet neutrino 
with high mass states $\nu_{\mu}$, $\nu_{\tau}$ is of the order 
\be
\sin^2 2 \theta_{\mu s} \sim \sin^2 2\theta_{es}\cdot 
\sin^2\theta  \sim 10^{-7} \ ,
\ee
so that the bound from the primordial nucleosynthesis 
\cite{nucl} can be  satisfied. 

The influence of the  
singlet fermion on ``standard" structure of the Zee
model  is negligibly small and the results of the previous 
sections are not changed. \\

\section{\bf  Conclusions}\par
\par
  1. Zee model reproduces rather  naturally the pattern of  
neutrino masses and mixing which  
solves the atmospheric neutrino problem,   
supplies a desired HDM component in the Universe and gives the 
$\bar \nu_{\mu} - \bar \nu_{e}$ oscillations 
in the range of sensitivity of  existing experiments. 

2. The solar neutrino problem can be solved in  extension 
of the model with an additional singlet fermion $s$, 
so that solar neutrinos undergo $\nu_e \rightarrow s$ conversion. 
The introduction of $s$ does not destroy  basic features of the 
Zee model. 

3. The  data on oscillations 
of solar and atmospheric neutrinos as well as the cosmological 
mass scale  
fix all parameters of the Zee mass matrix. The  scenario 
implies in general inverse flavour 
hierarchy of the Zee boson couplings. 
There is a  possibility of  
$f_{e \mu} \simeq f_{\mu \tau} \gg f_{e \tau}$
which may imply certain horizontal symmetry. 

4. The masses of  the charged scalar bosons are  
of the order 100 - 500 GeV,  and in certain cases 
at least one of the bosons 
can be as light as the  lower kinematical bound. 

5. The scenario will be tested in forthcoming experiments:\\ 
(i) the probability of  $\n_\m\A \n_\tau$ oscillations is 
expected to be very small; discovery of these oscillations 
in CHORUS and NOMAD will rule out the scenario.   
(ii) The signal of $\n_e\A \n_\tau$ oscillations 
may be in the region of sensitivity of these 
experiments.  
(iii) The confirmation of the LSND positive result 
will  further testify for the suggested scenario.  
(iv) One may expect deviations from the SM predictions 
in $\m\A \n_\m e \bar\n_e$ decay, 
$\bar \n_\m \  e^- \A \bar \n_\m \  e^-$ and  $\n_\m \  e^- \A  \n_\m
\  e^-$ scatterings, the violation of $e - \mu - \tau$ universality etc.. 
(v) The $\mu \A e\r$ decay can be close to the present experimental
upper bound. 
(vi) The  life time of  the neutrino
   radiative decay $\nu_{3(2)}\A \nu_1 \r$ is expected to be 
     $10^{22} - 10^{29}$ years. 
The decay of the relic neutrinos  may have 
observable astrophysical consequences.\\

\vskip 0.5cm

{\large {\bf Acknowledgments}} \par

 We would like to thank   Z. Hioki,  K. Hagiwara and 
K. Hikasa for the helpful discussion
  of radiative corrections.
  We also thanks   H. Shibuya for experimental informations of 
   $\n_e\A \n_\tau$ at CHORUS.
This research is supported by the Grant-in-Aid for Science Research,
Ministry of Education, Science and Culture, Japan(No. 07640413).


\newpage

\begin{center}
{\large{\bf Figure Captions}} \\
\end{center}
\noindent
{\bf Fig.1}\qquad
The dependence of the oscillation 
probabilities  $P(\n_e\A \n_\tau)$ 
at CHORUS and NOMAD (solid line), 
and  $P(\n_\m\A \n_e)$ at LSND (dashed line) 
on $\sin\th $ for   $\Delta m_{32}^2=10^{-2} \ \eV^2$ 
and $m_0=2.45 \ \eV$. \\

  \noindent
{\bf Fig.2} \qquad
The  ratios  $ f_{\mu \tau}/ f_{e \tau}$ (solid line) 
and $f_{e \mu}/ f_{e \tau}$ (dashed line)  
as the functions of $m_0$ for
$P(\n_\m\A \n_e)=1.5\times 10^{-3}$.
\\  

  \noindent
{\bf Fig.3} \qquad
   The lower bound for $\sin\phi$ as the function of  $M_{1}$
for    $P(\n_\m\A \n_e)$ = 
$3\times 10^{-3}$(solid curve), 
$3\times 10^{-4}$(long-dashed curve)
and $3\times 10^{-5}$(short-dashed curve).
For other parameters  we take  
(a) $\tan\b = 59.3$, $m_0=2.45\ \eV$, 
(b) $\tan\b = 2$,   $m_0=2.45 \ \eV$ 
(c) $\tan\b=59.3$,  $m_0=10 \ \eV$. 
In all the cases  $M_{2}=300 \ \G$.\\  
           
        \noindent 
{\bf Fig.4} \qquad
The dependence of the   branching ratio of 
$\mu \A e\r$ on $m_0$  
for  $\sin\phi=0.02$ (solid curve),  $ 0.05$ (dashed 
curve), and $0.1$(dashed-dotted curve). The values of other 
parameters are fixed as 
$\tan\b=20$, $M_{1}=500 \ \G$, $M_{2}=300 \ \G$
          and $\sin^2 2\th=2\times 10^{-3}$.
         The short vertical lines indicate  
 the lower bounds on $m_0$ from the muon decay. 
The experimental upper bound on 
 $B(\mu \A e\r)$   
is shown  by the horizontal dashed
        line.\\

\end{document}